\documentclass[aps,prl,twocolumn,amsmath,amssymb]{revtex4}
\usepackage{graphicx}% Include figure files
\begin{document}

\renewcommand{\a}{\hat{a}}
\newcommand{\ad}{\hat{a}^{\dagger}}
\renewcommand{\b}{\hat{b}}
\newcommand{\bd}{\hat{b}^{\dagger}}
\renewcommand{\S}{\hat{S}}
\newcommand{\tS}{\hat{\tilde{S}}}
\newcommand{\tg}{\tilde{g}}
\newcommand{\Q}{\hat{Q}}
\renewcommand{\Re}{{\rm Re}}
\renewcommand{\Im}{{\rm Im}}
\newcommand{\N}{\hat{N}}

\newcommand{\COMMENT}{\large\bf}
\newcommand{\FigWidth}{0.9\columnwidth}

\title{Molecular production at a wide
Feshbach resonance in Fermi-gas of cooled atoms.} %!!

\author{Deqiang~Sun}
\author{Ar.~Abanov}
\author{V.L.~Pokrovsky}
%\email[]{abanov@tamu.edu}
%\homepage[]{}
\affiliation{
            Department of Physics,
            MS 4242,
        Texas A\&M University,
            College Station, TX 77843-4242
}

\date{\today}
%\date{Last Change \LastChange}

\begin{abstract}
The problem of molecular production from degenerate gas of
fermions at a wide Feshbach resonance, in a single-mode
approximation, is reduced to the linear Landau-Zener problem for
operators. The strong interaction leads to significant
renormalization of the gap between adiabatic levels. In contrast
to static problem the close vicinity of exact resonance does not
play substantial role. Two main physical results of our theory is
the high sensitivity of molecular production to the initial value
of magnetic field and generation of a large BCS condensate
distributed over a broad range of momenta in inverse process of
the molecule dissociation. %!!
\end{abstract}

%\pacs{PACS numbers:75.10.Nr, 05.50.+q, 75.10.Jm}

\maketitle
%\tableofcontents

In recent years there have been numerous achievements in the area
of ultra-cold atomic physics. The major experimental tool for it
is the use of the Feshbach resonances (FR)
\cite{Stwalley1976,Tiesinga1993,
Inouye1998,Mies2000,Goral2004,Chwedenczuk2004,Timmermans1999},
which occurs when the energy of a quasibound molecular state
becomes equal to the energy of two free alkali atoms. The
magnetic-field dependence of the resonance allows precise tuning
of the atom-atom interaction strength in an ultracold
gas \cite{Stwalley1976}. Moreover, time-dependent magnetic fields
can be used to reversibly convert atom pairs into weakly bound
molecules \cite{Regal2003,Strecker2003,Cubizolles2003,Jochim2003,Donley2002,
Chin2003,Herbig2003,Durr2004}. This technique has proved to be
extremely effective in converting degenerate atomic gases of
fermions
\cite{Regal2003,Strecker2003,Cubizolles2003,Hodby2005,Greiner2003,
Jochim2003,Zwierlein2003} and bosons
\cite{Claussen2002,Herbig2003,Durr2004} into bosonic dimer
molecules.

The Feshbach resonance proceeds at sufficiently strong magnetic
fields so that electronic spins are polarized. The collisions in
s-channel of such Fermi-atoms is possible only if they have
different states of nuclear spins. In a typical
experiment \cite{Regal2003,Greiner2003} an admixture of atoms of
$\phantom{}^{40}K$ with the same total atomic spin $9/2$ but
different spin projection quantum numbers $-7/2$ and $-9/2$ was
used.

Theoretical works on the molecular production can be roughly
divided in two categories. The first is a phenomenology suggesting
that the pairs of molecules perform independently the Landau-Zener
(LZ) transitions \cite{Mies2000,Chwedenczuk2004}. Therefore the
total number of molecules in the end of the process is the LZ
transition probability multiplied by the number of pairs. The most
problematic in this approach is what should be accepted for the LZ
transition matrix element $\Delta$ (further we call it the LZ
gap). Direct calculation of the transition probability from a
microscopic Hamiltonian up to the 4-th order in the interaction
constant \cite{DP} shows that, in contrast to the assumption of
phenomenological works, the many-body effects are very essential.
Another category includes the works based on a simplified model
\cite{Timmermanns2001}, in which molecules have only one available
state mimicking the condensate
\cite{Tikhonenkov2006-OC,Javanainen2004,Williams2004}. Though the
numerical works of this category displayed a reasonable
temperature dependence, they could not give a clear physical
picture and detailed dependencies on parameters of the problem.
The series of semi-analytical works by Pazy {\it et al.}
\cite{Band2007,Tikhonenkov2006-OC,Pazy2005,Tikhonenkov2006} were
based on two contradicting assumptions as we show later.

In this letter we consider the process of molecule production from
fermi-gas of atoms after the FR is swept across the Fermi sea. The
accepted model is valid under assumption of strong interaction
equivalent to the condition of wide resonance \cite{Gurarie2007}.
We derive the closed equation for such process. We show that the
problem of molecular production from degenerate gas of fermions in
a single-mode approximation can be reduced to the linear LZ
problem for operators. In this respect our result agrees with the
conjecture of the phenomenological theories, but the strong
interaction leads to significant renormalization of the LZ gap,
which occurs to be independent on the fermi-gas density. Our
results display a significant dependence of the molecular
production on the initial state preparation. At the inverse
transformation of the BEC molecular gas into atomic gas the latter
appears in the state with strongly developed BCS condensate
directly after magnetic field sweeping.

Our starting point is the Timmermanns {\it et al.}
Hamiltonian \footnote{For simplicity, we focus on a homogeneous
system of volume $V$.}
\begin{eqnarray}
&&\hat{H}=\sum_{p,\sigma}(\epsilon(t)+\epsilon_p)\ad_{p\sigma}\a_{p\sigma}
+\sum_{p}\omega_{p}\bd_{p}\b_{p}+\nonumber\\
&&\frac{g}{\sqrt{V}}\sum_{p,q}\left(\b_{q}\ad_{p+q\uparrow}\ad_{-p\downarrow}+
\bd_{q}\a_{-p\downarrow}\a_{p+q\uparrow}\right) \label{eq:Ham}
\end{eqnarray}
describing fermionic atoms, created by $\ad_{p\sigma}$ with
momentum $p$, “spin” $\sigma=\uparrow,\downarrow$ which
distinguishes between the two internal states of the atoms, and
kinetic energy $\epsilon_{p}$ , that are coupled to diatomic
bosonic  molecules with kinetic energy $\omega_q$ created by
$\bd_{q}$. The position and the width (molecular lifetime) of the
FR are respectively controlled by the bare time-dependent detuning
energy $\epsilon(t)$ and the coupling constant $g$, the former
experimentally tunable by magnetic field. The Hamiltonian % !!
\eqref{eq:Ham} neglects nonresonant atom-atom and
molecule-molecule interactions that near a FR are subdominant to
the resonant scattering.

In this letter we use the word "atoms" for both, uncoupled atoms
and those inside molecules, and the word "fermions" for atoms that
are not bound in molecules. Correspondingly, we denote the number
of atoms $\N$ and the number of fermions $\N_{F}$.

First we demonstrate that the condition of wide resonance allows
to neglect kinetic energy of fermions. Indeed, the calculations in
weak coupling approximation \cite{DP} show that the characteristic
interaction energy is $g\sqrt{n}$, where $n=N/V$ is the density of
atoms. If $g\sqrt{n}$ exceeds the Fermi energy of fermions
$\epsilon_{F}=\hbar^{2}(3\pi^{2}n)^{2/3}/2m$ the fermion % !!
dispersion can be neglected. This requirement is equivalent to:
\begin{equation}\label{eq:Gamma}
\Gamma =\frac{(2mg/\hbar^{2})^{2}} {(3\pi^{2})^{4/3}} n^{-1/3}\gg
1
\end{equation}
which coincides with the definition of the wide resonance
\cite{Gurarie2007}. The same criterion allows to neglect the
dispersion of molecules. This approximation has been used earlier
\cite{Barankov2004,Pazy2005} and numerical proof of its
validity has been reported in \cite{Tikhonenkov2006}.

In what follows we assume that only molecules with zero momentum
$\bd_{q=0}$ are produced, neglecting molecules $\bd_{q\not=0}$
excited above the molecular condensate (the single-mode
approximation). This assumption was justified for the equilibrium
state in the case of wide resonance \cite{Gurarie2007} and used in
the most of theoretical works on the dynamics of transition
\cite{Barankov2004, Goral2004, Pazy2005, Tikhonenkov2006,
Band2007}. We thus replace $\b_q$ in Eq. \eqref{eq:Ham} by
$\b_0\delta_{q,0} \equiv \b \delta_{q,0}$. With this
simplification one notices that only a pair of fermions with
opposite momenta and spin can be converted into a molecule. The
two assumptions allow to simplify the Hamiltonian
\eqref{eq:Ham} and to solve the problem exactly. %!!

Let us introduce spin operators $\S_z,\S_{-}$ and $\S_{+}=\S_{-}^{\dagger}$ as follows: %!!
\begin{equation}
\S_z=\frac{1}{2}\sum_p\left(\hat{n}_{p\uparrow}+\hat{n}_{-p\downarrow}-1\right)\qquad
\S_{-}=\sum_p\a_{p\uparrow}\a_{-p\downarrow}; \label{eq:Spins}
\end{equation}
with the standard commutation relations $[\S_{+},\S_{-}]=2\S_z$,
$[\S_z,\S_{+}]=\S_{+}$. Neglecting fermion and Boson kinetic
energy one can rewrite the Hamiltonian \eqref{eq:Ham} in terms of
spin operators:
\begin{equation}
\hat{H}=2\epsilon(t)\S_z+\frac{g}{\sqrt{V}}\left(\b\S_{+}+\bd\S_{-}\right)
\label{eq:HamS}
\end{equation}
This Hamiltonian commutes with the operator $\Q =\S_z+\bd\b$
equivalent to the total number of atoms $N$ and additionally with
the square of total spin operator ${\bf
S}^2=S_z^2+S_{+}S_{-}-S_z$. It follows from the first equation
\eqref{eq:Spins} that $S_{z}=(\N_{F}-N_{s})/2$, where $\N_{F}$ is
the number of fermions and $N_{s}$ is the number of available
fermionic states. Thus, $\Q=(\N - N_s)/2$. Since $N_{s}\gg
N\geqslant N_{F}$ we can approximate $S_z \approx -N_{s}/2$.

The Heisenberg equations of motion are:
\begin{equation}
\hbar\dot{\b}=-i\tg\S_{-};\qquad %!!
\hbar\dot{\S}_{-}=-2i\epsilon(t)\S_{-}+2i\tg\bd\S_z %!!
\label{eq:motionZ}
\end{equation}
Generally these equations are non-linear. However, in the
wide-resonance approximation $S_z = -N_{s}/2$, they become linear.
Eliminating $S_{-}$, we arrive at an ordinary linear differential
equation for the operator $\b$:

\begin{equation}
\hbar^2\ddot{\b}+2i\hbar\epsilon(t)\dot{\b}+\Delta^2\b=0 %!!
\label{c-equation}
\end{equation}
where $\Delta =g\sqrt{n_s}$ and $n_{s}=N_{s}/V$ is the density of
available states. Linear equation \eqref{c-equation} turns into
the parabolic cylinder equation if $\epsilon (t) $ is a linear
function of time. In the LZ theory it describes the evolution of
the amplitude to find the system in one of its two states. The
role of the LZ gap is played by the value $\Delta =g\sqrt{n_{s}}$,
which strongly exceeds $g\sqrt{n}$. The characteristic time during
which the LZ transition takes place $\tau _{LZ}$ is determined by
requirement that the instantaneous frequency $2\epsilon ( \tau
_{LZ}) $ becomes equal to $\Delta $. Thus, the characteristic
value of $\epsilon (t)$ is equal to $\Delta =g\sqrt{n_{s}}\gg
\epsilon _{F}$.

The model neglecting the dispersion is valid until the kinetic
energy $\frac{p^{2}}{2m}\ll g\sqrt{n_{s}}$. The value $p_{s}$
limiting available states is determined by equation:
\begin{equation}
\frac{p_{s}^{2}}{2m}=g\sqrt{n_{s}}  \label{ps-ns}
\end{equation}
and the density of available states $n_s$ is associated with the
limiting momentum $p_s$ by a standard relation:
\begin{equation}
n_{s}=\frac{(p_{s}/\hbar )^{3}}{3\pi ^{2}}=
\frac{(2mg/\hbar^2)^{6}}{(3\pi^2)^4} =n\Gamma^{3};\text{ }
\label{ns-delta}
\end{equation}
Both $n_s$ and $\Delta$ are independent on the density of fermions
$n$. The condition of the wide resonance \eqref{eq:Gamma} requires
that $N_{s}\gg N$ thus justifying the accepted approximation. In a
series of works by Pazy, Tikhonenkov {\it et al.}
\cite{Pazy2005,Tikhonenkov2006}, the authors neglected
dispersion and put $N_s=N$. As it was demonstrated above the
neglecting of dispersion is justified only if the condition of the
wide resonance \eqref{eq:Gamma} is satisfied. But the same
condition ensures that $N_s\gg N$. Thus, their two fundamental
assumptions physically contradict each other, though their model
is mathematically consistent. Exact quantum solution of the same %!!
problem was recently found by Altland and Gurarie 
\footnote{Private communication.}. %!!

The value $g$ can be extracted from the experimental data on the
dependence of the scattering length on magnetic field near the FR
\cite{Regal2004} using a well-known relation:
$g=\hbar\sqrt{4\Pi(a-a_0)\varepsilon /m}$. On the other hand it
can be estimated theoretically as $g\sim \epsilon
_{hf}\sqrt{a_{0}^{3}}$, where $\epsilon _{hf}$ is the hyperfine
energy and $a_{0}$ is the radius of the molecule far from the
resonance. Both these estimates give for $^{40}$K $g\sim
10^{-28}erg\times cm^{3/2}$ and $\Delta =0.006K$. The sweeping of
magnetic field proceeds with amplitude few hundred Gauss. It
corresponds to energy scale about $0.03K$, larger than $\Delta $.

The first important conclusion is that the dynamic LZ problem is
simpler than the static equilibrium one. In the latter problem
most interesting and intriguing phenomena take place at very small
$\epsilon \leq \varepsilon_{F}$. In this range of energy the
s-scattering amplitude reaches its unitary limit and scattering
length changes sign resulting in strong BCS coupling and complex
behavior of the atomic and molecular densities. It is not the case
in the dynamic problem since the effective interval for
transformations of atoms into molecules is $\Delta \gg \varepsilon
_{F}$. This statement is correct for any sweeping rate above relaxation.

Another important conclusion is that the strong interaction leads
to renormalization of the LZ gap. The energy scale which appears
in the perturbation theory is $\Delta ^{\left( 0\right)
}=g\sqrt{n}$ \cite{DP}. In the case of wide resonance
$\Delta=g\sqrt{n_s} $ is much larger than $\Delta ^{(0)}$ and does
not depend on the density of atoms.

Employing equations \eqref{eq:motionZ}, the general solution of
the ordinary differential equation \eqref{c-equation} reads:
\begin{eqnarray}
\b\left( t\right) =u\left( t,t_{0}\right) \b\left( t_{0}\right) -
i\tg v\left(t,t_{0}\right) \S_{-}\left( t_{0}\right),&&  \label{general}\\
i\tg\S_{-}(t)=-\dot{u}\left( t,t_{0}\right)\b\left( t_{0}\right)+
i\tg \dot{v}\left(t,t_{0}\right) \S_{-}\left( t_{0}\right),&&
\label{generalS}
\end{eqnarray}
where $u\left( t,t_{0}\right) $ and $v\left( t,t_{0}\right) $ are
standard solutions of the same equation satisfying initial
conditions $u\left( t_{0},t_{0}\right) =1$, $\dot{u}\left(
t_{0},t_{0}\right) =0$ and $v\left( t_{0},t_{0}\right) =0$,
$\dot{v}\left( t_{0},t_{0}\right) =1$. These solutions have the
following properties:
\begin{equation}\label{eq:prop}
\left\vert u\right\vert ^{2}+ \Delta^{-2}\left\vert
\dot{u}^{2}\right\vert = \Delta ^{2}\left\vert v^{2}\right\vert +
\left\vert \dot{v}^{2}\right\vert =1;
\dot{u}^{\ast}\dot{v}+\Delta^2u^{\ast}v=0
\end{equation}

The solution (\ref{general},\ref{generalS}) allows to follow the
evolution of the number of molecules $N_m(t)=\langle \bd\b
\rangle(t)$, the BCS condensate amplitude $F(t)$ defined by
equation $F^2(t)=\langle \S_{+}\S_{-}\rangle(t)$, and the BCS-BEC
coherence factor $C(t)=\langle \bd \S_{-}\rangle(t)$. In the case
when initial coherence factor is zero it is given by:
\begin{eqnarray}
&&N_m(t)=|u|^2N_m(t_0)+\tg^2|v|^2F^2(t_0)\label{eq:A}\\
&&\tg^2F^2(t)=|\dot{u}|^2N_m(t_0)+\tg^2 |\dot{v}|^2F^2(t_0)
\label{eq:B}\\
\!\!\!\!&&C(t)=i\tg^{-1}\dot{u}u^{\ast}N_m(t_0)+i\dot{v}v^{\ast}F^2(t_0),
\label{eq:C}
\end{eqnarray}
where $\tg =g/\sqrt{V}$. Using \eqref{eq:prop} and summing eqs.
\eqref{eq:A} and \eqref{eq:B} we find that
\begin{eqnarray}\label{eq:const1}
N_sN_m(t)+F^2(t)=\mbox{const},
\end{eqnarray}
which is a consequence of the conservation laws. Since for any
state $F^2(t)>0$, if there are no molecules in the initial state, %!!
their number $N_m(t)$ can not exceed a value $ F^2(t_{0})/N_{s}$
at any time. Therefore, the final molecule production rate depends %!!
on the initial state. Below we consider two experimentally most
relevant situations: no molecules and only molecules and no
fermions in the initial state. In both these cases the initial
value $C=\langle \bd \S_{-}\rangle(t_{0})=0$.

In the case of no molecules in the initial state
$N_m(t_{0})=0$ general equations %!!
(\ref{eq:A},\ref{eq:C}) simplify to
\begin{equation}\label{average-t}
N_m(t)= \tg^2 |v|^{2}F^2(t_0);\qquad i\tg C(t)=
-\tg^2\dot{v}v^{\ast}F^2(t_0)
\end{equation}
Evolution of $F(t)$ in this case is completely determined by
\eqref{eq:const1}. Note, that the coherence factor $C(t)$ does not
remain zero. If at the initial moment there exists both condensate
of molecules $\langle \b\rangle $ and the BCS condensate $\langle
\S_{-}\rangle$, their time evolution can be obtained from
equations (\ref{general},\ref{generalS}) by taking averages from
both sides.

The value $F^2(t_0)$ strongly depends on the initial state. We
consider first an initial state defined as the filled Fermi
sphere. The condensate density can be readily found for this
state: $F^2= N/2$. Thus, the final average number of molecules in
this case is very small $N_m(+\infty)\leq N/(2N_s)$. To produce a
reasonable fraction of molecules it is necessary to have large
condensate amplitude in the initial state. A natural way to
generate such initial state is to start with sufficiently small
negative detuning energy $\epsilon$. The effective dimensionless
BCS coupling constant reads $\lambda_{BCS}=-\nu_F g^2/\epsilon$,
where $\nu_F$ is the density of state at the Fermi-energy
\cite{Gurarie2007}. This constant becomes of the order of 1 at
$\mid\epsilon\mid\sim\Gamma\epsilon_F$, a value of detuning
between Fermi-energy and the gap $\Delta$. When detuning becomes
less than this value the condensate spreads from an exponentially
narrow spherical layer near the Fermi sphere to a sphere of the
radius given by the cut-off momentum $p_c$. The latter is inverse
proportional to the effective radius of the hyperfine interaction
and is much larger than $p_s$. Therefore, the molecular production
determined by the part of initial BCS condensate confined in the
sphere of the radius $p_s$ first grows and then decreases when the
initial detuning decreases. 

Such a strong dependence of the final
molecular production on the initial state, in particular on the
value of the initial magnetic field explains why different
experimenters obtain different fractions of molecules in the final
state even in the adiabatic regime  
\cite{Strecker2003, Donley2002, Zwierlein2003, Greiner2003}.
%\cite{hulet, wieman, ketterle,jin}. 
Note, that in experiments in which a significant molecular
production was achieved the initial state was indeed close to the
FR, whereas the final state could be sufficiently far from it.

Thus, in a realistic experimental setup the initial value of
$\epsilon $ is small $\epsilon_0\leq\Gamma\epsilon_F\ll \Delta$
and then $\epsilon$ increases linearly with time. In this case one
can put $t_0=0$, and $\epsilon (t) =\dot{\epsilon}t$.  Equation
\eqref{c-equation} turns into the parabolic cylinder equation. Its
standard solution $u(t,0)$ has an asymptotic property:
$|u(\infty,0)|^2=\exp (-\pi \Delta^{2}/2\hbar\dot{\epsilon})$. %!!
Employing it together with \eqref{eq:prop} and
\eqref{average-t}, we arrive at the following number of
molecules in the final state:
\begin{equation}
N_m (+\infty )=F^2_0N_{s}^{-1}\left[ 1-\exp \left( -\pi \Delta
^{2}/2\hbar\dot{\epsilon}\right) \right] \label{LZ-answer-bb} %!!
\end{equation}
It can be proven that the maximal possible value of $F^2$ is
$N_sN/2$. It corresponds to complete transformation of atoms into
molecules in the adiabatic regime $\dot{\epsilon}\rightarrow 0$.
Equation \eqref{LZ-answer-bb} looks exactly as the LZ transition
probability multiplied by an effective number of pairs, but in
contrast to phenomenological theories
\cite{Mies2000,Chwedenczuk2004} and the perturbation theory result
\cite{DP} the coefficient in front of $1/\dot{\epsilon}$ in
exponent does not depend on the initial density of the atoms. This
theoretical prediction can be checked experimentally.

Finally, we consider an inverse process with no fermions, no BCS
condensate and only the molecular condensate in the initial state:
$\langle \b \rangle ( -\infty ) =\sqrt{N/2}$ and sweeping of the
magnetic field in opposite direction. Then in the end the
condensate density is determined by the LZ value: $\langle \b
\rangle ( +\infty ) = \sqrt{N/2}\exp(-\pi \Delta ^{2}/
2\hbar\dot{\epsilon})$, whereas the absolute value of the BCS
condensate amplitude $\langle\S_{-}\rangle $ can be found from the
conservation law \eqref{eq:const1}:
\begin{equation}
\vert \langle \S_{-}\rangle \vert ^{2}=\frac{N_{s}N}{2} \left[
1-\exp \left( -\frac{\pi \Delta ^{2}}{\hbar\dot{\epsilon}}\right) %!!
\right]   \label{bcs}
\end{equation}
This result has a clear physical interpretation. It corresponds to
$\frac{N}{2}\left[ 1-\exp \left( -\pi \Delta
^{2}/\hbar\dot{\epsilon} \right) \right] $ Cooper pairs %!!
distributed with identical probability $w=\frac{N}{2N_{s}}\left[
1-\exp \left( -\pi \Delta ^{2}/\hbar\dot{\epsilon}\right) \right]$ %!!
between $N_{s}$ available states. Then the modulus of the pair
amplitude at a fixed state is $\langle
\a_{\mathbf{p}\uparrow}\a_{-\mathbf{p}\downarrow}\rangle
=\sqrt{w}$. If all these amplitudes have the same phase, the total
condensate amplitude is equal to $\langle\S_{-} \rangle
=N_{s}\sqrt{w}$, which is equivalent to equation \eqref{bcs}. This
result has experimentally verifiable consequence. Indeed, the pair %!!
created after the sweeping of magnetic field has the size in real
space $r_{pair}=\frac{\hbar }{p_{s}}\ll n^{-1/3}$. It means that
the pair is a compact formation well separated from
other pairs and, therefore, can be rather qualified as a %!!
quasimolecule. In contrast to real molecules the quasimolecules
have parallel electron spins. The second their peculiarity is
their instability: after the sweeping of magnetic field stops,
they decay during the relaxation time. The latter is rather long
since the pair fermion collisions do not produce energy
relaxation. The experimental estimate for the relaxation time is
in the range of seconds. Therefore, it seems quite feasible to
switch off the trap for much shorter time and observe the
correlations of momenta and spins in runaway particles. The
prediction of our theory is that the correlation prefers opposite
velocities and parallel spins in the range of energy up to $\Delta
$.

In conclusion, we considered the molecule formation and
dissociation in cooled Fermi-gas when magnetic field is swept
across the wide Feshbach resonance. It was demonstrated that in
this situation the fermion kinetic energy is negligible. The
resulting molecular production from initial fermions is described
by LZ-like formulae with strongly renormalized LZ gap independent
on the density of initial fermions. However, the molecular
production strongly depends on initial value of magnetic field. In
the inverse process of molecular dissociation immediately after
the sweeping stops, there appear Cooper pairs with parallel
electronic spins and opposite momenta homogeneously distributed
within a sphere in the momentum space, whose radius $p_s$ is much
larger than the Fermi momentum. Another experimentally verifiable
prediction is independence of the coefficient in front of
$1/\dot{\epsilon}$ in LZ exponents entering the equations for
molecular production \eqref{LZ-answer-bb} and the BCS condensate
amplitude \eqref{bcs} on the initial density of atoms (molecules).

We acknowledge illuminating discussions with Victor Gurarie and
Leo Radzyhovsky. V.P. is thankful to the Kavli Institute of
Theoretical Physics, UCSB, for the hospitality extended to him
during the Workshop on coherence effects in cold gases and
strongly correlated systems, May - June 2007. This work was partly
supported by the DOE under the grant DE-FG02-06ER46278 and KITP
NSF Grant No. PHY05-51164.

%\begin{figure}
%\includegraphics[scale=1.2]{WZNW}
%\includegraphics[width=\FigWidth]{figures/Scut}
%\caption{\label{fig:Scut} \dots
%}
%\end{figure}
%\bibliography{MolecularProduction}
%\bibliographystyle{prl}
%\begin{thebibliography}{99}

%\bibitem{}

%\end{thebibliography}
\bibliography{MolecularProduction-Short}

\end{document}